 \documentclass[12pt]{iopart}
 
\usepackage{iopams}  
\usepackage{graphics}
\usepackage{epsfig}
\usepackage{xcolor}
 
\begin{document}
{Letter to the Editor}	

 	\title[Correct scaling of the correlation length for concentrated electrolytes]{Correct scaling of the correlation length from a theory for concentrated electrolytes}

 	\author{A Ciach$^1$ and O Patsahan$^2$}
 	
\address{$^1$Institute of Physical Chemistry, Polish Academy of Sciences, 01-224 Warszawa, Poland }
 \address{$^2$Institute for Condensed Matter Physics, National Academy of Sciences of Ukraine, Lviv, Ukraine }
 
 	\vspace{10pt}
% 	\begin{indented}
% 		\item[]August 2017
% 	\end{indented}
 
 \begin{abstract}
 Self-consistent theory for concentrated electrolytes is developed.  Oscillatory decay of the charge-charge correlation function with the decay length  that shows perfect agreement with the experimentally discovered and so far unexplained scaling  is obtained.   
 For the density-density correlations, monotonic asymptotic decay with the decay length comparable with the decay length of the charge correlations is found. We show that the correlation lengths in concentrated electrolytes depend crucially on the local variance of the charge density. 
 \end{abstract}
\vspace{2pc}
\submitto{\JPCM}

 \maketitle
 
The solvation force acting on the plates confining electrolytes decays with increasing distance between the plates with the decay length $\lambda_S$ equal to the correlation length in the bulk. The results of experimental measurements for dilute electrolytes confirm that the decay length is equal to the Debye screening length, $\lambda_D$,
 in perfect agreement with theoretical predictions.
 $\lambda_D$  decreases with  increasing  concentration of ions, $\rho$. As shown in experiments by Perkin and coauthors~\cite{smith:16:0,lee:17:0}, however,  $\lambda_S$ in concentrated electrolytes  {\it increases} with increasing $\rho$ {\it instead of further  decreasing}, in contrast to theoretical predictions of   classical theories~\cite{hansen:06:0}. Moreover, $\lambda_S$  follows the scaling relationship $\lambda_S/\lambda_D\sim(a/\lambda_D)^3$, where  $a$ is the ion diameter \cite{lee:17:0}. This relationship was verified for simple salts in water, ionic liquid solutions and alkali halide solutions. Support for  the long  decay
 length has been recently provided by using an independent technique in~\cite{Gaddam2019}. 

 Several theoretical attempts have been made to tackle the problem. In particular, an explanation of the long decay length has been proposed within the framework of the dressed ion theory extension~\cite{Kjellander2018,Kjellander2019}.
 The scaling law for the screening length has been confirmed
in \cite{Goodwin2017,Ludwig2018,Rotenberg_2018,Adar2019,deSouza2020} based on different assumptions, introducing e.g., short-range repulsive and attractive non-Coulomb interactions between the ions, and/or  solvent structure or some modification of the short-range part of the Coulomb potential. However, the scaling exponents found in these studies as well as in all-atom molecular dynamics simulations \cite{Coles2020,Zeman2020} appeared to be significantly lower than the experimentally measured one. Furthermore, no evidence for
an anomalously long-ranged, monotonic decay in effective ionic
interactions revealed in the experiments has been found in the very recent all-atom molecular dynamics simulations \cite{Zeman2020}. The puzzle remains unsolved.
 
 In this letter we consider both, the charge-charge and the density-density correlation functions in the mesoscopic theory developed for ionic systems and for mixtures in \cite{ciach:05:0,ciach:11:2}. 
 In this theory,
 we consider dimensionless charge and number density of ions in mesoscopic regions around ${\bf r}$,   $c({\bf r})=\rho_+({\bf r})-\rho_-({\bf r})$ and
   $\rho({\bf r})=\rho_+({\bf r})+\rho_-({\bf r})$ respectively.  $\pi\rho_i({\bf r})/6$ is the fraction of the volume 
  of the mesoscopic region that is covered by the ions of the $i$-type, with $i=+,-$. The average diameter of the ions and solvent molecules is denoted by $a$, and serves as a length unit. The above fields 
   can be considered as constraints imposed on the microscopic states. The grand potential for fixed $c({\bf r})$ and  $\rho({\bf r})$ takes the standard form
   \[
    \Omega_{co}[c,\rho]=
   U_{co}[c,\rho]-TS -\mu\int d{\bf r}\rho({\bf r}),
   \]
where
  \[
    U_{co}[c,\rho]=
    \frac{1}{2}\int d{\bf r}_1\int d{\bf r}\frac{e^2c({\bf r}_1)c({\bf r}_1+{\bf r})\theta(|{\bf r}|-1)}{\epsilon |{\bf r}|}
   \]
is the internal energy due to the electrostatic interactions, $T$ is temperature, $\mu$ is the chemical potential of the 
ions, and $S[c,\rho]$ 
is the entropy in the presence of the constraints $c({\bf r})$ and $\rho({\bf r})$.  We make the approximation
$-TS=\int d{\bf r} f_h(c({\bf r}),\rho({\bf r}))$, where $f_h(c,\rho)$ is the free-energy density 
corresponding to the entropy of mixing of ions and solvent,
\[
\beta f_h(c,\rho)=\rho_+\ln\rho_++\rho_-\ln\rho_-+(\rho_{tot}-\rho)\ln (\rho_{tot}-\rho),
\]
where $\beta=1/k_BT$, with $k_B$ the Boltzmann constant.
For simplicity we assume $\rho_{tot}=1$.
All lengths are dimensionless (in $a$-units). The Heaviside $\theta$ function prevents from
contributions to the internal energy from overlapping hard cores of the ions.
The subscript $co$ indicates that the microscopic states incompatible with the constraints  $c({\bf r})$ and $\rho({\bf r})$ 
do not contribute to $ \Omega_{co}$.  When the constraints $c({\bf r})$ and $\rho({\bf r})$ are released, such microscopic states, i.e. fluctuations $\phi$ and $\eta$ around the fields $c$ and $\rho$ can appear. The average charge and number densities can remain equal to  $c({\bf r})$ and $\rho({\bf r})$, if the fluctuations cancel one another, i.e. $\langle \phi\rangle=\langle \eta\rangle=0$. Note, however that these additional microscopic states give additional contribution to the grand potential.

$c=0$ in the absence of electrostatic field independently of the concentration of ions. What distinguishes dilute and concentrated electrolytes is the variance of the charge density, $\langle \phi^2({\bf r})\rangle$. Imagine a window at ${\bf r}$ with the size comparable with the size of ions. In the course of time, solvent molecules and ions enter and leave the window. 
The charge in the window vanishes when it is occupied by the solvent, but  when a cation or an anion enters the window, it becomes positively or negatively charged.
It happens the more frequently the larger is the concentration of the ions. The average deviation of the local charge from zero is independent of ${\bf r}$, and is given by $\sqrt{\langle \phi^2\rangle}$. 

In order to see the role of $\langle \phi^2\rangle$ for highly concentrated electrolytes,
let us 
divide the system into windows described above, and assume that  the charge in each window is either $+\sqrt{\langle \phi^2\rangle}$ or $-\sqrt{\langle \phi^2\rangle}$. 
Entropy favors random distribution of the $+$ and $-$ signs among the cells, but oppositely charged nearest neighbors are favored by the energy. For two windows separated by the distance $\Delta r$, there are 4 possible combinations of the $+,-$ signs. When the charge is correlated so that the energy of this pair is negative, $+,-$ and $-,+$ signs are left. The excess free energy in $k_BT$ units associated with fixing opposite charges is roughly $\beta \Delta F= -l_B\langle \phi^2\rangle/\Delta r+\ln 2$, where $l_B=\beta e^2/\epsilon$ is the Bjerrum length in $a$ units. $\beta \Delta F<0$ for $\Delta r<l_B\langle \phi^2\rangle/\ln 2$, and a rough estimation for the correlation length is $\sim l_B\langle \phi^2\rangle$. Note that  $\langle \phi^2\rangle\sim\rho$, since the variance of a fluctuating quantity should be proportional to the number of fluctuating objects. Thus, the correlation length is $\lambda_S/a\sim l_B \rho$. This scaling is equivalent to $\lambda_S/\lambda_D\sim(a/\lambda_D)^3$ because $a/\lambda_D=\sqrt{4\pi l_B\rho}$, and is therefore in  agreement with  experiments~\cite{lee:17:0,smith:16:0,lee2017}.
The above considerations  highlight the significance of the variance of the charge density, and a need for a first-principles theory that takes it into account.

Fluctuations $\phi$ and $\eta$ 
appear with the probability proportional to $\exp(-\beta \Delta\Omega_{co})$, where $\Delta\Omega_{co}$ is the excess grand potential associated with appearance of the considered fluctuations~\cite{landau}.  Taking into account the fluctuation contribution, we obtain 
 \begin{equation}
  \label{F}
   \beta \Omega[ c,\rho]=\beta\Omega_{co}[c,\rho]-\ln \int D\eta\int D\phi \exp\Big[-\beta \Delta\Omega_{co}[c,\rho;\phi,\eta]\Big].
  \end{equation}
  The leading-order contributions to $\Omega$ coming from the fluctuation term in (\ref{F}) are proportional to $\langle \phi^2\rangle$ when periodic $\phi({\bf r})$ minimizes the energy ~\cite{ciach:18:0,patsahan:21:0}.
The correlation functions  $G_{cc}(r)=\langle \phi({\bf r}_1)\phi({\bf r}_1+{\bf r})\rangle$ and  $G_{\rho\rho}(r)=\langle \eta({\bf r}_1)\eta({\bf r}_1+{\bf r})\rangle$ can be calculated with the probability distribution proportional to $\exp(-\beta \Delta\Omega_{co})$. 
 On the other hand, $G_{cc}$ and $G_{\rho\rho}$ are inverse to the second functional derivative of $ \beta \Omega[ c,\rho]$ with respect to the corresponding fields. 
 $G_{c\rho}=0$ for $c=0$.
 
 We make the self-consistent Gaussian approximation 
 \[
\beta \Delta\Omega_{co}[c,\rho;\phi,\eta]\approx \beta H_G[c,\rho;\phi,\eta]
\]
  with
  \[ 
  \beta H_G[c,\rho;\phi,\eta]=\frac{1}{2}\int d{\bf r}_1\int d{\bf r}_2
   \Big[ \phi({\bf r}_1)C_{cc}(r) \phi({\bf r}_2)+\eta({\bf r}_1)C_{\rho\rho}(r) \eta({\bf r}_2)
 %  +2 \phi({\bf r}_1)C_{c\rho}(r)\eta({\bf r}_2)
   \Big],
  \]
  where $C_{cc}$ and $C_{\rho\rho}$ are the second functional derivatives of $\beta\Omega$ with respect to $c$ and $\rho$, respectively.  $H_G$ incorporates  higher-order terms in the expansion of $\beta \Delta\Omega_{co}$, when the fluctuation contribution is present in (\ref{F}).
 With this approximation,  both ways of calculating  the  correlation functions lead to the same results. 
 
 When calculating the functional derivatives of $\Omega$ 
 up to the second order,  we take into account the fluctuation contribution in (\ref{F}), but neglect it for  higher-order derivatives, and  
 obtain in Fourier representation 
\begin{eqnarray}
\label{Ccc3}
 \tilde C_{cc}(k)= l_B\frac{4\pi\cos(k)}{k^2} +
 A_{0,2}+\frac{A_{0,4}}{2}\langle\phi^2\rangle +\frac{A_{2,2}}{2}\langle\eta^2\rangle 
 \nonumber\\
 -A_{1,2}^2\int d{\bf r}e^{i{\bf k}\cdot{\bf r}}G_{cc}(r)G_{\rho\rho}(r)
\end{eqnarray}
\begin{eqnarray}
\label{Crr3}
 \tilde C_{\rho\rho}(k)=
 A_{2,0}+\frac{A_{2,2}}{2}\langle\phi^2\rangle +\frac{A_{4,0}}{2}\langle\eta^2\rangle
 -\frac{A_{1,2}^2}{2}\int d{\bf r}e^{i{\bf k}\cdot{\bf r}}G_{cc}(r)^2
  \nonumber\\
 -\frac{A_{3,0}^2}{2}\int d{\bf r}e^{i{\bf k}\cdot{\bf r}}G_{\rho\rho}(r)^2
\end{eqnarray}
 where   $\tilde C_{cc}(k)=1/\tilde G_{cc}(k)$,  $\tilde C_{\rho\rho}(k)=1/\tilde G_{\rho\rho}(k)$ and 
\begin{equation}
\label{Amn}
 A_{m,n}(c,\rho)=\frac{\partial^{n+m}(\beta f_h)}{\partial^n c\partial^m \rho}.
\end{equation}

Note that both, $\int d{\bf r}e^{i{\bf k}\cdot{\bf r}}G_{cc}(r)^2$ and $\int d{\bf r}e^{i{\bf k}\cdot{\bf r}}G_{\rho\rho}(r)^2$ take a maximum for $k=0$, and for $k\to 0$, (\ref{Crr3}) takes the form
$\tilde C_{\rho\rho}(k)=R_0+R_2k^2+...$.
From the $k\to 0$ form of $\tilde C_{\rho\rho}(k)$, we obtain the asymptotic decay of correlations in the real space 
\begin{equation}
\label{Grras}
\label{Grr}
 G_{\rho\rho}(r)=A_{\rho}\exp(-r/\xi_{\rho})/r,
\end{equation}
where $4\pi A_{\rho}\xi_{\rho}^2R_0=1$ and $4\pi A_{\rho}R_2=1$.

Let us focus on  concentrated electrolytes where charge waves with the wavenumber $k_0\sim \pi$ (oppositely charged nearest neighbors) are energetically favored. 
A theory for a fluctuating field $\phi$ with the lowest energy assumed for  $\phi$ that oscillates in space with the  wavenumber $k_0>0$  was developed by Brazovskii~\cite{brazovskii:75:0}.  
 He noted that a local variance of an oscillatory field is large, because the field is typically either larger or smaller from its average value. Thus,  $\langle\phi^2\rangle=(2\pi)^{-3}\int d{\bf k}/\tilde C_{cc}(k)$
should be taken into account, but  $\langle\eta^2\rangle$ can be neglected. Using the above assumptions and (\ref{Amn}) we obtain
\begin{eqnarray}
\label{Crr4}
 \tilde C_{\rho\rho}(k)\approx
 \frac{1}{\rho(1-\rho)}+\frac{1}{\rho^3} \int \frac{d{\bf k}}{(2\pi)^3\tilde C_{cc}(k)}
 -\frac{1}{2\rho^4}\int d{\bf r}e^{i{\bf k}\cdot{\bf r}}G_{cc}(r)^2
 \nonumber\\
-\frac{(2\rho-1)^2}{2\rho^4(1-\rho)^4}\int d{\bf r}e^{i{\bf k}\cdot{\bf r}}G_{\rho\rho}(r)^2.
\end{eqnarray}
Note that for $\rho\to 1$ the first and the last terms in (\ref{Crr4}) diverge. Neglecting the remaining terms and using (\ref{Grras}), we obtain the result, valid in the asymptotic regime
\[
\xi_{\rho}\simeq_{\rho\to 1} \frac{1.54}{(1-\rho)^{1/3}}
\hspace{10mm}
\rm{and} 
\hspace{10mm}
%\label{Ar}
A_{\rho}\simeq_{\rho\to 1} 0.44 (1-\rho)^{5/3}.
\]

Let us consider  $\tilde C_{cc}(k)$ and focus on the last term in (\ref{Ccc3}). From the charge neutrality it follows that $\int d{\bf r}G_{cc}(r)=0$. If $\xi_{\rho}$ is large, $\int d{\bf r}G_{cc}(r)G_{\rho\rho}(r)$ is small, because $G_{\rho\rho}(r)$ is almost constant for $r$ such that $G_{cc}(r)$ differs significantly from zero. 
 Based on the above observation, we assume that the last term in (\ref{Ccc3}) can be neglected, and  obtain
\begin{equation}
\label{Ccc4}
 \tilde C_{cc}(k)\approx l_B\frac{4\pi\cos(k)}{k^2} +
\frac{1}{\rho}+\frac{1}{\rho^3}\int \frac{d{\bf k}}{(2\pi)^3\tilde C_{cc}(k)}.
\end{equation}
On the  large density side of the Kirkwood line~\cite{kirkwood:36:0,leote:94:0,ciach:03:1} (oscillatory decay of $G_{cc}(r)$),
(\ref{Ccc4}) can be solved analytically in the Brazovskii approximation, 
when $\beta \tilde V(k)=l_B4\pi\cos(k)/k^2$ is Taylor-expanded about its minimum at $k=k_0\approx 2.46$, and the expansion is truncated. We take into account that $\tilde V(k)$ is an even function of $k$, and make the approximation   $\beta \tilde V(k)\approx \beta \tilde V(k_0)+\beta v (k^2-k_0^2)^2$, where $\beta v \approx 0.044 l_B$. For small $\tilde C_{cc}(k_0)$ we have the Brazovskii result~\cite{brazovskii:75:0},  
\begin{equation}
\label{<phi2>}
 \langle \phi^2\rangle=\int \frac{d{\bf k}}{(2\pi)^{3}[\tilde C_{cc}(k_0)+\beta v (k^2-k_0^2)^2]}\approx \frac{k_0}{4\pi\sqrt{\tilde C_{cc}(k_0)\beta v}},
\end{equation}
 where $\tilde C_{cc}(k_0)$ is the self-consistent solution of (\ref{Ccc4}) with $k=k_0$ and (\ref{<phi2>}), and its lengthy explicit  expression is given in \cite{ciach:12:0,ciach:18:0}.
In this approximation   
\begin{equation}
\label{Gccr}
G_{cc}(r)\approx \frac{A_c\sin(k_0 r)\exp(-\alpha_0 r)}{r}, 
\end{equation}
with
\begin{equation}
\label{al0}
 A_c\approx \langle \phi^2\rangle/k_0 \hskip1cm {\rm and} \hskip1cm\alpha_0^{-1}\approx8\pi \beta v\langle \phi^2\rangle\approx 1.1l_B \langle \phi^2\rangle.
\end{equation}
 The analytical solution was obtained under the assumption of small $\tilde C_{cc}(k_0)$, therefore  (\ref{al0}) 
 can be valid only for   $\tilde C_{cc}(k_0)<1$. Both $\rho$ and $l_B$ are large for  $\tilde C_{cc}(k_0)<1$, and only for such parameters our results are presented. 

\begin{figure}[h]
 \centering
\includegraphics[scale=0.33]{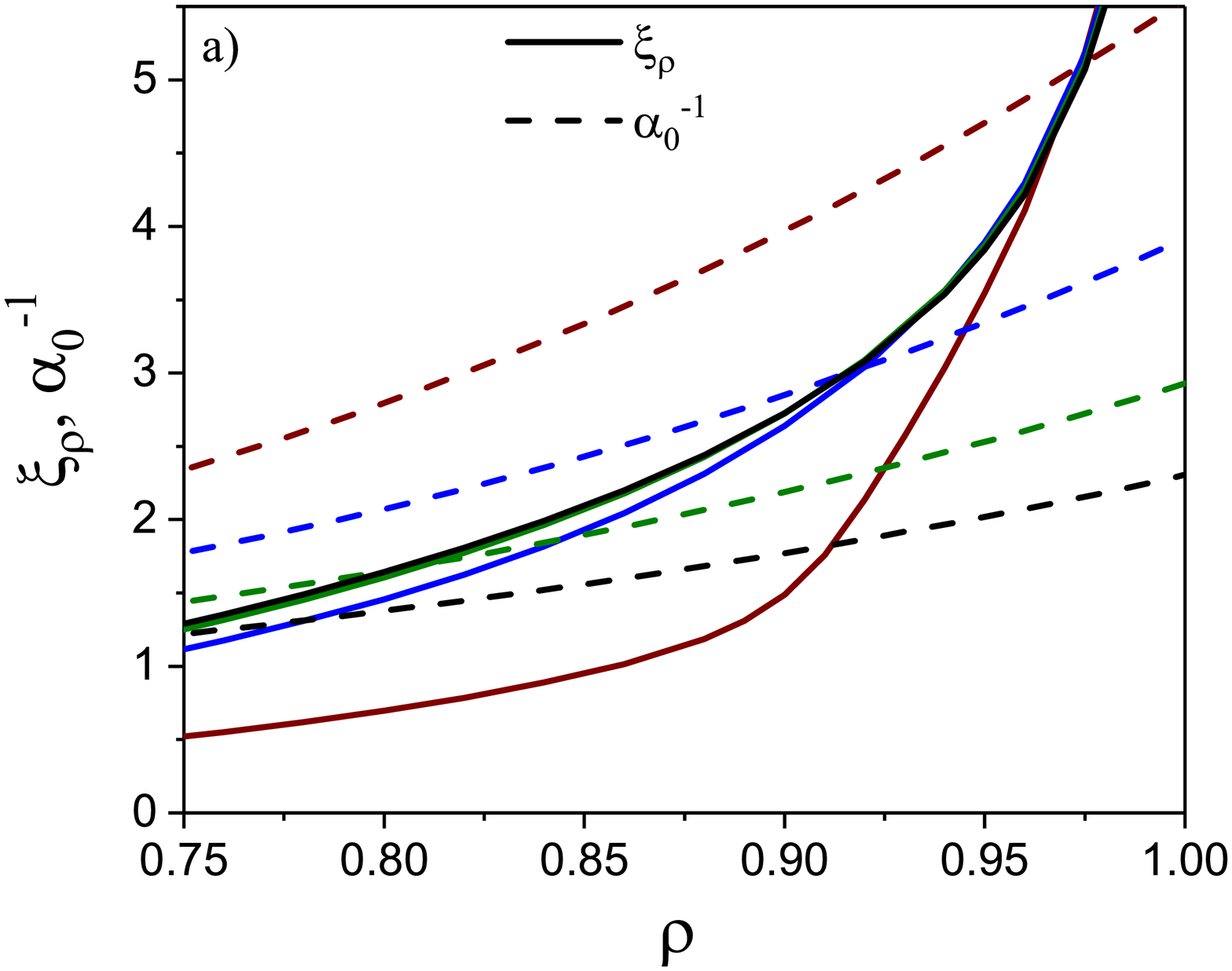}
\includegraphics[scale=0.33]{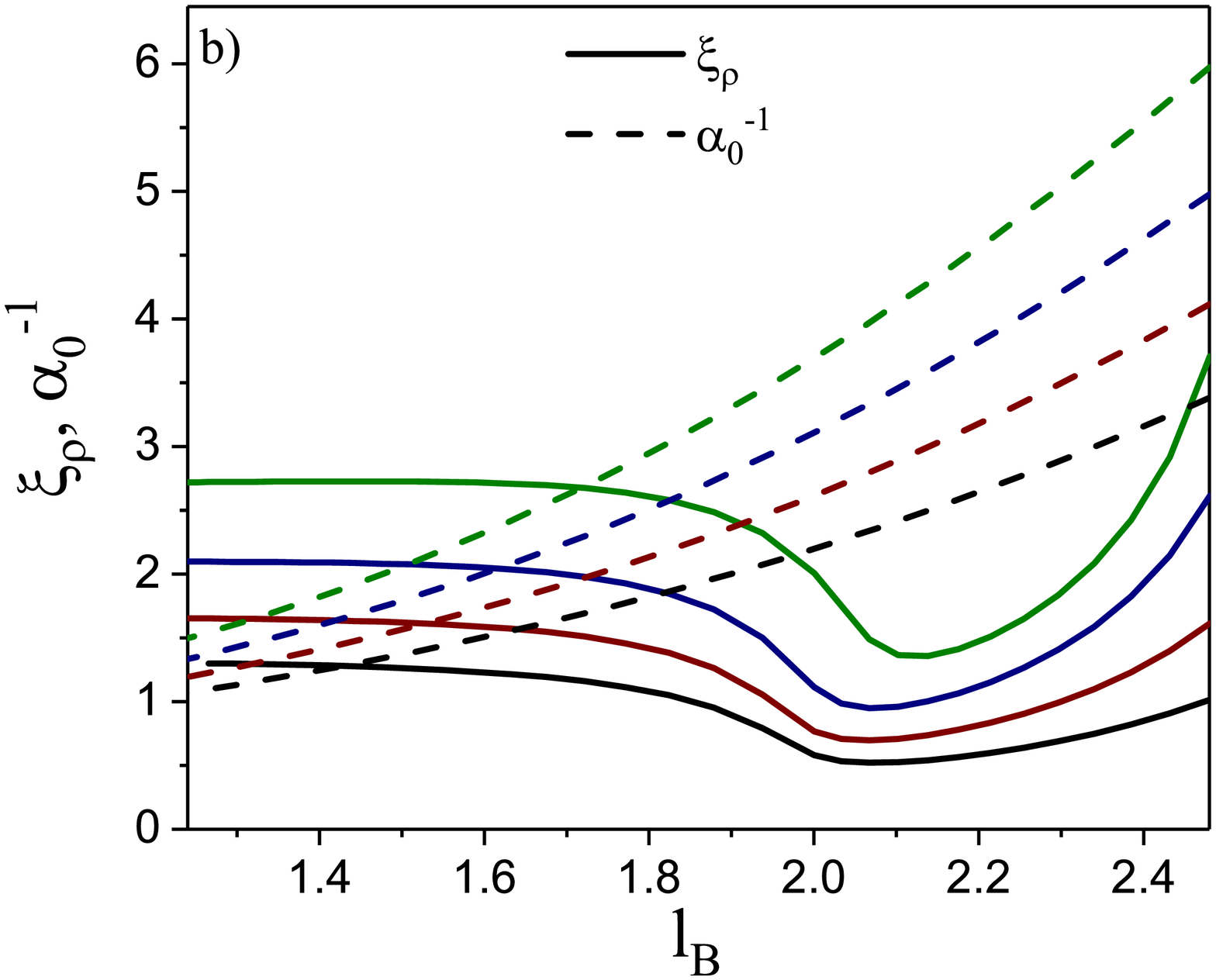}
\caption{ The correlation length of the density-density correlations (solid lines) and of the charge-charge correlations (dashed lines) as a function of $\rho$ for  $l_B=1.38,1.55,1.77, 2.07$, black, green, blue and red lines, respectively (panel~a) and  as a function of $l_B$ for  $\rho=0.75, 0.8, 0.85, 0.9$  from the bottom to the top line,  respectively (panel~b).  $\rho$ denotes the mole fraction of the ions, and $l_B=\beta e^2/\epsilon$ is the Bjerrum length. $l_B,\xi_{\rho}$ and $1/\alpha_0$ are in units of the average molecular size $a$. }
\label{f1}
\end{figure}

\begin{figure}[h]
 \centering
\includegraphics[scale=0.33]{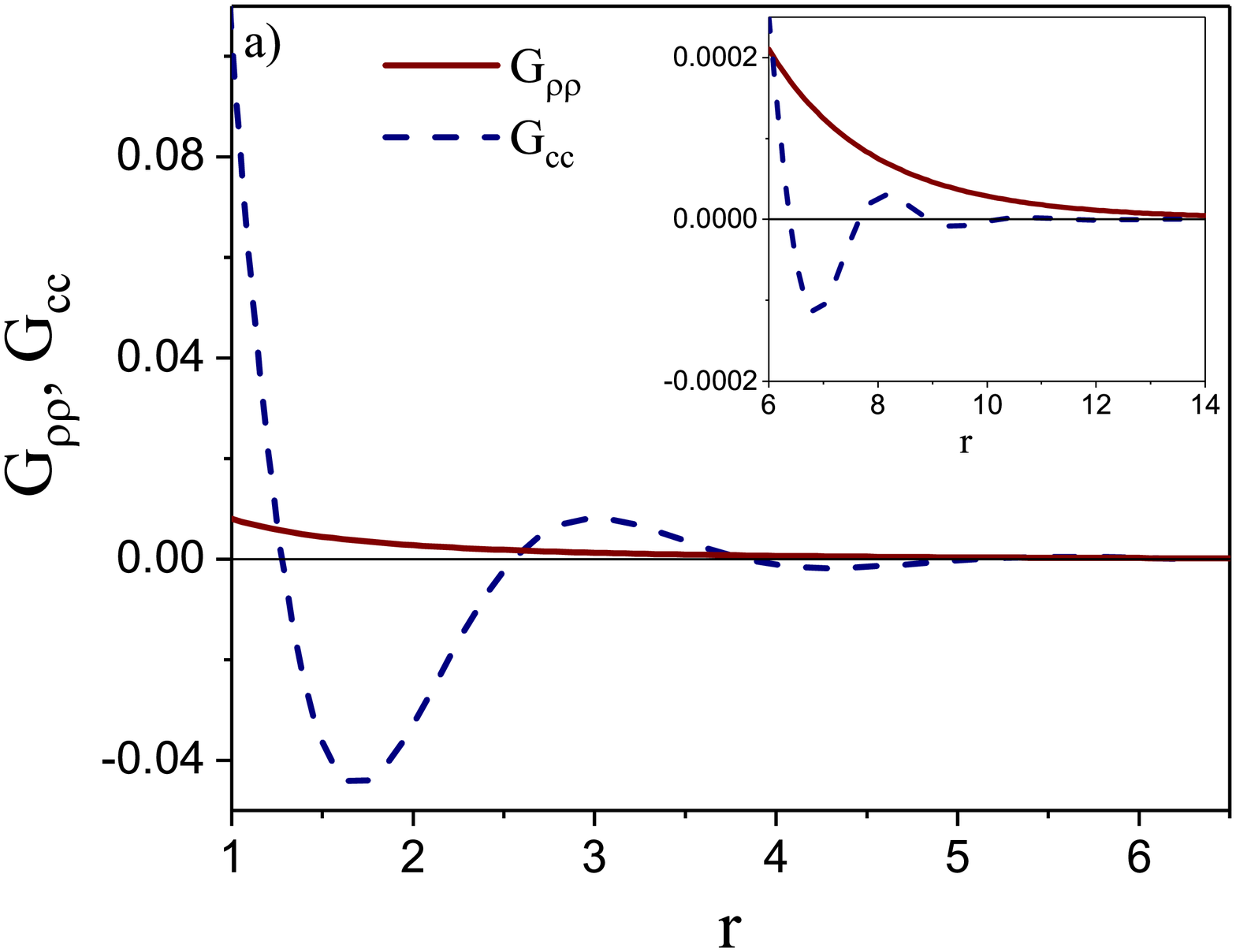}
\includegraphics[scale=0.33]{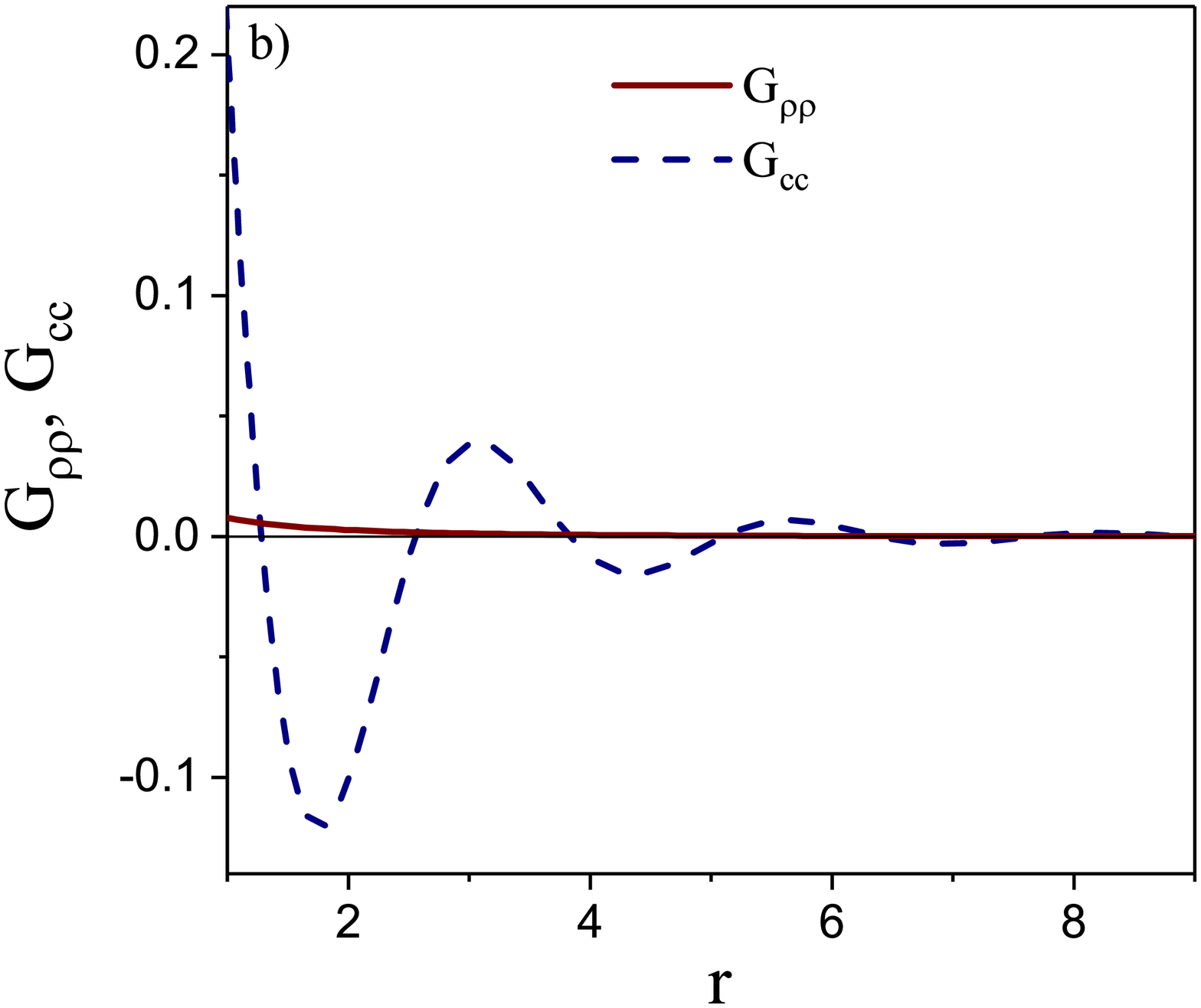}
\caption{ The density-density (solid line) and the charge-charge (dashed line) correlation functions. Panel a:   $\rho=0.9$ and  $l_B=1$. Note that the oscillatory decay of charge correlations dominates for $r<6$, and the monotonic decay of density correlations dominates for $r>6$ (see the inset). Panel~b: the same correlation functions but for $\rho=0.9$ and $l_B=1.55$.}
\label{f3}
\end{figure}

(\ref{Gccr})-(\ref{al0}) allow us  to compute  $\xi_{\rho}$ and $A_{\rho}$ for intermediate $\rho$.
In figure~\ref{f1}a, the dependence of the decay lengths of the  correlation functions on $\rho$ is shown for several values of  $l_B$ ($1.38\le l_B\le 2.07$). In figure~\ref{f1}b, the decay lengths as functions of $l_B$ are shown for  several values of $\rho$ ($0.75\le\rho\le 0.9$). In figure~\ref{f3}, the density-density and the charge-charge correlation functions are shown for $\rho=0.9$,  and for $l_B=1$ (panel~a) and $l_B=1.55$ (panel~b).

 $\xi_{\rho}$ increases for increasing $\rho$ in a strongly nonlinear way, and its dependence on $l_B$ is nonmonotonic. For sufficiently small $l_B$ (high temperature) and large $\rho$, $\xi_{\rho}> 1/\alpha_0$, i.e. the number density fluctuations are correlated over larger distances than the charge fluctuations. Moreover, $G_{\rho\rho}(r)$ decays monotonically. We should stress, however that $G_{\rho\rho}$ depends strongly on the assumed form of the entropy associated with the packing effects, and is therefore model (and experimental system) dependent.
 $G_{cc}$, in contrast, depends only on the entropy associated with mixing, and our result for this function should be more universal. As shown in figure~\ref{f1},
 $1/\alpha_0>\xi_{\rho}$  for large $l_B$. 
 Notably, $1/\alpha_0$ has a nearly linear  dependence  on $\rho$ for fixed $l_B$, with the slope increasing with $l_B$, and a nearly linear  dependence on $l_B$ for fixed $\rho$, with a slope increasing with $\rho$, indicating that $\alpha_0^{-1}\sim l_B \rho$, in agreement with experiments.
 This result and (\ref{al0}) confirm the conjecture $\langle \phi^2\rangle \propto \rho$ and show the  crucial role of the variance of the charge density, as already discussed on the heuristic level.   
 
 Our approximate theory is not expected to predict accurately the  constant $C$ in the relation $\alpha_0^{-1}/l_B=C\rho$. $C$ depends significantly on the assumed average diameter of the ions in the experimental relation $\rho=a^3 c_{ion}$, and may also depend on the other microscopic details. As follows from our rough estimate, however, it is of the same order of magnitude in the theory and experiment.

The  solvation force between plates confining the dense electrolyte exhibits damped oscillations at short distances, and a monotonic decay at large separations~\cite{smith:16:0}. We obtain the correct scaling of the correlation length of the charge-charge correlations, which, however, decay in an oscillatory way. A monotonic decay  is found for the density-density  correlation function, but its decay length does not obey the scaling.
 The two decay lengths may be close to each other in real systems, and the decay of the solvation force requires a separate study.

 {\bf Acknowledgments}
 
 We thank R. Evans, M. Holovko and S. Kondrat for discussions.
  \vspace{5mm}
  \providecommand{\newblock}{}

%\bibliographystyle{iopart-num} 
%\bibliography{bibliography_20.bib}
 \end{document}